\documentclass[nofootinbib,superscriptaddress, aps, prl, 10pt, amsmath, amssymb, bibnotes,
altaffilletter, twocolumn, floatfix]{revtex4-2}

\usepackage{graphicx}
\usepackage{dcolumn}
\usepackage{bm}
\DeclareUnicodeCharacter{2212}{-}

\usepackage{hyperref}
\usepackage[T1]{fontenc}
\usepackage{lineno}
\usepackage{xcolor}
\usepackage{soul}

\graphicspath{{figures/}}

\hypersetup{colorlinks=true, citecolor=blue, urlcolor=blue, linkcolor=blue}
\usepackage{lineno}

\renewcommand{\sc}{\textsc}

\def\DEM{\sc{Demonstrator}}
\def\MJD{\sc{Majorana Demonstrator}}
\def\onbb{$0\nu\beta\beta$}

\def\gag{$g_{a\gamma}$}
\def\gae{$g_{ae}$}

\def\enrge{${}^{\mathrm{enr}}$Ge}
\def\natge{${}^{\mathrm{nat}}$Ge}

\begin{document}


\title{Search for solar axions via axion-photon coupling with the \textsc{Majorana Demonstrator}}

\newcommand{\uw}{Center for Experimental Nuclear Physics and Astrophysics, and Department of Physics, University of Washington, Seattle, WA 98195, USA}
\newcommand{\ITEP}{National Research Center ``Kurchatov Institute'' Institute for Theoretical and Experimental Physics, Moscow, 117218 Russia}
\newcommand{\JINR}{Joint Institute for Nuclear Research, Dubna, 141980 Russia} 
\newcommand{\lbnl}{Nuclear Science Division, Lawrence Berkeley National Laboratory, Berkeley, CA 94720, USA}
\newcommand{\lbnle}{Engineering Division, Lawrence Berkeley National Laboratory, Berkeley, CA 94720, USA}
\newcommand{\lanl}{Los Alamos National Laboratory, Los Alamos, NM 87545, USA}
\newcommand{\queens}{Department of Physics, Engineering Physics and Astronomy, Queen's University, Kingston, ON K7L 3N6, Canada}
\newcommand{\unc}{Department of Physics and Astronomy, University of North Carolina, Chapel Hill, NC 27514, USA}
\newcommand{\duke}{Department of Physics, Duke University, Durham, NC 27708, USA}
\newcommand{\ncsu}{Department of Physics, North Carolina State University, Raleigh, NC 27695, USA}	
\newcommand{\ornl}{Oak Ridge National Laboratory, Oak Ridge, TN 37830, USA}
\newcommand{\ou}{Research Center for Nuclear Physics, Osaka University, Ibaraki, Osaka 567-0047, Japan}
\newcommand{\pnnl}{Pacific Northwest National Laboratory, Richland, WA 99354, USA}
\newcommand{\ttu}{Tennessee Tech University, Cookeville, TN 38505, USA}
\newcommand{\sdsmt}{South Dakota Mines, Rapid City, SD 57701, USA}
\newcommand{\usc}{Department of Physics and Astronomy, University of South Carolina, Columbia, SC 29208, USA}
\newcommand{\usd}{Department of Physics, University of South Dakota, Vermillion, SD 57069, USA}  
\newcommand{\ut}{Department of Physics and Astronomy, University of Tennessee, Knoxville, TN 37916, USA}
\newcommand{\tunl}{Triangle Universities Nuclear Laboratory, Durham, NC 27708, USA}
\newcommand{\mpi}{Max-Planck-Institut f\"{u}r Physik, M\"{u}nchen, 80805, Germany}
\newcommand{\tum}{Physik Department and Excellence Cluster Universe, Technische Universit\"{a}t, M\"{u}nchen, 85748 Germany}
\newcommand{\williams}{Physics Department, Williams College, Williamstown, MA 01267, USA}
\newcommand{\ciemat}{Centro de Investigaciones Energ\'{e}ticas, Medioambientales y Tecnol\'{o}gicas, CIEMAT 28040, Madrid, Spain}
\newcommand{\iu}{Department of Physics, Indiana University, Bloomington, IN 47405, USA}
\newcommand{\iuceem}{IU Center for Exploration of Energy and Matter, Bloomington, IN 47408, USA}

\author{I.J.~Arnquist}\affiliation{\pnnl} 
\author{F.T.~Avignone~III}\affiliation{\usc}\affiliation{\ornl}
\author{A.S.~Barabash}\affiliation{\ITEP}
\author{C.J.~Barton}\affiliation{\usd}	
\author{K.H.~Bhimani}\affiliation{\unc}\affiliation{\tunl} 
\author{E.~Blalock}\affiliation{\ncsu}\affiliation{\tunl} 
\author{B.~Bos}\affiliation{\unc}\affiliation{\tunl} 
\author{M.~Busch}\affiliation{\duke}\affiliation{\tunl}	
\author{M.~Buuck}\altaffiliation{Present address: SLAC National Accelerator Laboratory, Menlo Park, CA 94025, USA}\affiliation{\uw} 
\author{T.S.~Caldwell}\affiliation{\unc}\affiliation{\tunl}	
\author{Y-D.~Chan}\affiliation{\lbnl}
\author{C.D.~Christofferson}\affiliation{\sdsmt} 
\author{P.-H.~Chu}\affiliation{\lanl} 
\author{M.L.~Clark}\affiliation{\unc}\affiliation{\tunl} 
\author{C.~Cuesta}\affiliation{\ciemat}	
\author{J.A.~Detwiler}\affiliation{\uw}	
\author{Yu.~Efremenko}\affiliation{\ut}\affiliation{\ornl}
\author{H.~Ejiri}\affiliation{\ou}
\author{S.R.~Elliott}\affiliation{\lanl}
\author{G.K.~Giovanetti}\affiliation{\williams}  
\author{M.P.~Green}\affiliation{\ncsu}\affiliation{\tunl}\affiliation{\ornl}   
\author{J.~Gruszko}\affiliation{\unc}\affiliation{\tunl} 
\author{I.S.~Guinn}\affiliation{\unc}\affiliation{\tunl} 
\author{V.E.~Guiseppe}\affiliation{\ornl}	
\author{C.R.~Haufe}\affiliation{\unc}\affiliation{\tunl}	
\author{R.~Henning}\affiliation{\unc}\affiliation{\tunl}
\author{D.~Hervas~Aguilar}\affiliation{\unc}\affiliation{\tunl} 
\author{E.W.~Hoppe}\affiliation{\pnnl}
\author{A.~Hostiuc}\affiliation{\uw} 
\author{M.F.~Kidd}\affiliation{\ttu}	
\author{I.~Kim}\affiliation{\lanl}
\author{R.T.~Kouzes}\affiliation{\pnnl}
\author{T.E.~Lannen~V}\affiliation{\usc} 
\author{A.~Li}\affiliation{\unc}\affiliation{\tunl} 
\author{A.M.~Lopez}\affiliation{\ut}	
\author{J.M. L\'opez-Casta\~no}\affiliation{\ornl} 
\author{E.L.~Martin}\altaffiliation{Present address: Duke University, Durham, NC 27708, USA}\affiliation{\unc}\affiliation{\tunl}	
\author{R.D.~Martin}\affiliation{\queens}	
\author{R.~Massarczyk}\affiliation{\lanl}		
\author{S.J.~Meijer}\affiliation{\lanl}	
\author{T.K.~Oli}\affiliation{\usd}  
\author{G.~Othman}\altaffiliation{Present address: Universit{\"a}t Hamburg, Institut f{\"u}r Experimentalphysik, Hamburg, Germany}\affiliation{\unc}\affiliation{\tunl}
\author{L.S.~Paudel}\affiliation{\usd} 
\author{W.~Pettus}\affiliation{\iu}\affiliation{\iuceem}	
\author{A.W.P.~Poon}\affiliation{\lbnl}
\author{D.C.~Radford}\affiliation{\ornl}
\author{A.L.~Reine}\affiliation{\unc}\affiliation{\tunl}	
\author{K.~Rielage}\affiliation{\lanl}
\author{N.W.~Ruof}\affiliation{\uw}	
\author{D.C.~Schaper}\affiliation{\lanl} 
\author{D.~Tedeschi}\affiliation{\usc}		
\author{R.L.~Varner}\affiliation{\ornl}  
\author{S.~Vasilyev}\affiliation{\JINR}	
\author{J.F.~Wilkerson}\affiliation{\unc}\affiliation{\tunl}\affiliation{\ornl}    
\author{C.~Wiseman}\affiliation{\uw}
\author{W.~Xu}~\email{wenqin.xu@usd.edu}\affiliation{\usd} 
\author{C.-H.~Yu}\affiliation{\ornl}
\author{B.X.~Zhu}~\email{xiaoyu.x.zhu@jpl.nasa.gov}\altaffiliation{Present address: Jet Propulsion Laboratory, California Institute of Technology, Pasadena, CA 91109, USA}\affiliation{\lanl}

\collaboration{{\sc{Majorana}} Collaboration}
\noaffiliation

\date{\today}

\begin{abstract}
Axions were originally proposed to explain the strong-CP problem in QCD. Through the axion-photon coupling, the Sun could be a major source of axions, which could be measured in solid state detection experiments with enhancements due to coherent Primakoff-Bragg scattering. The \MJD\ experiment has searched for solar axions with a set of $^{76}$Ge-enriched high purity germanium detectors using a 33 kg-yr exposure collected between Jan. 2017 and Nov. 2019. A temporal-energy analysis gives a new limit on the axion-photon coupling as \gag$<1.45\times 10^{-9}$ GeV$^{-1}$ (95\% confidence level) for axions with mass up to 100 eV/$c^2$. This improves laboratory-based limits between about 1 eV/$c^2$ and 100 eV/$c^2$.
\end{abstract}


\maketitle

Axions were originally motivated as the Peccei-Quinn solution to the strong-charge parity (CP) problem in quantum chromodynamics (QCD)~\cite{Peccei1977, Peccei1977a, Weinberg1978, Wilczek1978a, Peccei2008}, where a CP-violating parameter in the strong force is heavily constrained by neutron electric dipole moment measurements~\cite{Baker2006, Kim2010, Pendlebury2015}. 
Axion models have evolved from the original QCD axion characterized by a constrained relationship between mass and coupling constant to a group of pseudoscalar particles beyond the Standard Model (SM), called axion-like particles (ALPs). Axions and ALPs are dark matter candidates~\cite{Preskill1983, Abbott1983, Dine1983} and can serve as a portal to the dark sector~\cite{SNOWMASS13}, and they motivate active global searches with a comprehensive range of techniques~\cite{Asztalos2005,Kim2010, Rosenberg2015, Graham2015, Irastorza2018}.
Axions can couple to photons directly and via neutral pions, so the axion-photon coupling \gag\ is typically nonzero~\cite{Sikivie2021}. 
In the Kim-Shifman-Vainshtein-Zakharov (KSVZ) generic hadronic model~\cite{Kim1979, Shifman1980}, axions are considered ``invisible,'' with no tree-level axion-electron coupling \gae; however, KSVZ axions couple to photons as expressed in Eq.~\ref{eq:ksvz}, where $N$ and $E$ are the dimensionless coefficients for color and electromagnetic anomalies. A window of $0.25\leq|E/N - 1.92| \leq 12.75$ has been determined for realistic QCD KSVZ axions~\cite{DiLuzio2017}. 
\begin{equation}
    \label{eq:ksvz}
    g_{a\gamma}=\frac{m_a}{\rm{eV}}\frac{2.0}{10^{10}\rm{GeV}}\bigg(\frac{E}{N}-1.92 \bigg)
\end{equation} 

The axion-photon coupling \gag\ in electromagnetic fields enables the Primakoff effect~\cite{Primakoff1951}, where axions can convert into photons, and its inverse process.
Astrophysical and cosmological measurements such as stellar cooling rates have set stringent limits on \gag~\cite{Dicus1978, Raffelt1986, Raffelt1988, Kim2010}.
A recent analysis of horizontal branch (HB) stars in a sample of 39 galactic globular clusters gives the limit \gag$ < 0.66 \times 10^{-10}$ GeV$^{-1}$ at a 95\% confidence level (CL), one of the strongest constraints to date in a wide mass range~\cite{Ayala2014}.
Complementary to these observations, laboratory-based axion searches provide valuable independent tests with a variety of physics and experimental techniques~\cite{Raffelt2008, Graham2015, Irastorza2018, Sikivie2021}, probing a large phase space of QCD models.

The Sun may be a major source of axions~\cite{Raffelt1986, Pospelov2008, Gondolo2009, Derbin2011, Redondo2013}, producing axions by Primakoff scattering of thermal photons in the electromagnetic field of the solar plasma (the inverse Primakoff effect)~\cite{Dicus1978, Raffelt1986, Raffelt1988}.
A modern parameterization of the solar axion flux from the Primakoff process is given by  Eq.~\ref{eq:flux_new}, where $E_a$ is the axion energy in keV, $\lambda \equiv ($\gag$\times10^8~\rm{GeV})^4$, and $\Phi_0~\equiv~6.02~\times~10^{14}$~cm$^{-2}$~sec$^{-1}$~keV$^{-1}$~\cite{Bahcall2004, Andriamonje2007, Raffelt2008, Anastassopoulos2017}.
\begin{equation}
    \label{eq:flux_new}
    \frac{d\Phi}{dE_a}=\sqrt\lambda\Phi_0{E_a}^{2.481}e^{-{E_a}/{1.205}}
\end{equation} 

In the presence of strong magnetic fields, solar axions can be converted to electromagnetic signals. This mechanism, the Primakoff effect, is the one used on Earth by axion helioscope experiments. The CERN Axion Solar Telescope (CAST) experiment tracks solar axions over a wide range of mass using a dipole magnet originally built for the Large Hadron Collider and its best limit is similar to HB limits for $m_a \lesssim 0.02$ eV/$c^2$~\cite{Andriamonje2007, Arik2015, Anastassopoulos2017}. 
However, the CAST experiment quickly loses its sensitivity at the threshold of $m_a \sim 1.2$ eV/$c^2$, where solid state detectors may begin to play a significant role. 

As solar axions reach detector materials, they may also convert into photons by the intense local electric field of the target atomic nucleus. High Purity Germanium (HPGe) detectors and other solid state detectors may enjoy a great enhancement to these axion-induced photon signals when the condition of coherent Bragg diffraction is satisfied. The Bragg condition correlates photon energies $E$ with axion incident angles in respect to crystalline planes of the detectors, producing a strong dependence of the axion-photon conversion probability on the Sun's position at a given time $t$ and the crystal plane orientation. This is a unique signature of solar axions with both time- and energy-dependent features~\cite{Buchmuller1990, Paschos1994, Creswick1998}.

Pioneered by the SOLAX~\cite{Avignone1998} experiment, a series of solid state experiments probed \gag~using this coherent Primakoff-Bragg scattering technique~\cite{Avignone1998, Bernabei2001, Morales2002, Ahmed2009, Armengaud2013}.
The current best limits of these experiments, \gag\ $ < 1.7\times 10^{-9}$ GeV$^{-1}$ at 90\% CL and \gag $<2.15 \times 10^{-9}$ GeV$^{-1}$ at 95\% CL, were established by DAMA NaI detectors~\cite{Bernabei2001} and EDELWEISS-II germanium detectors~\cite{Armengaud2013}, respectively. As the photon energies are assumed to be the same as the kinetic energies of incident axions, these limits are valid for an axion mass from near zero up to 100 eV/$c^2$, where the mass energy of axions begin to yield a non-negligible energy correction.

The [001] axis of a Ge crystal is determined by the crystal boule axis, which corresponds to the vertical direction in most experiments, and only the horizontal orientation of the crystal planes (the azimuthal angle $\phi$) needs to be measured to predict solar axion signatures at a given time.
This approach was used by CDMS for their germanium detectors with a few-degree precision~\cite{Ahmed2009}.
Predictions for the energy- and time-dependent signals from coherent Primakoff-Bragg axion signatures were given in Refs.~\cite{Avignone1998, Cebrian1999, Avignone2009} and explored specifically for HPGe detectors at the location of the Sanford Underground Research Facility (SURF) in Lead, SD in Ref.~\cite{Xu2017}.
 
 \begin{figure}
    \includegraphics[width=8.3cm]{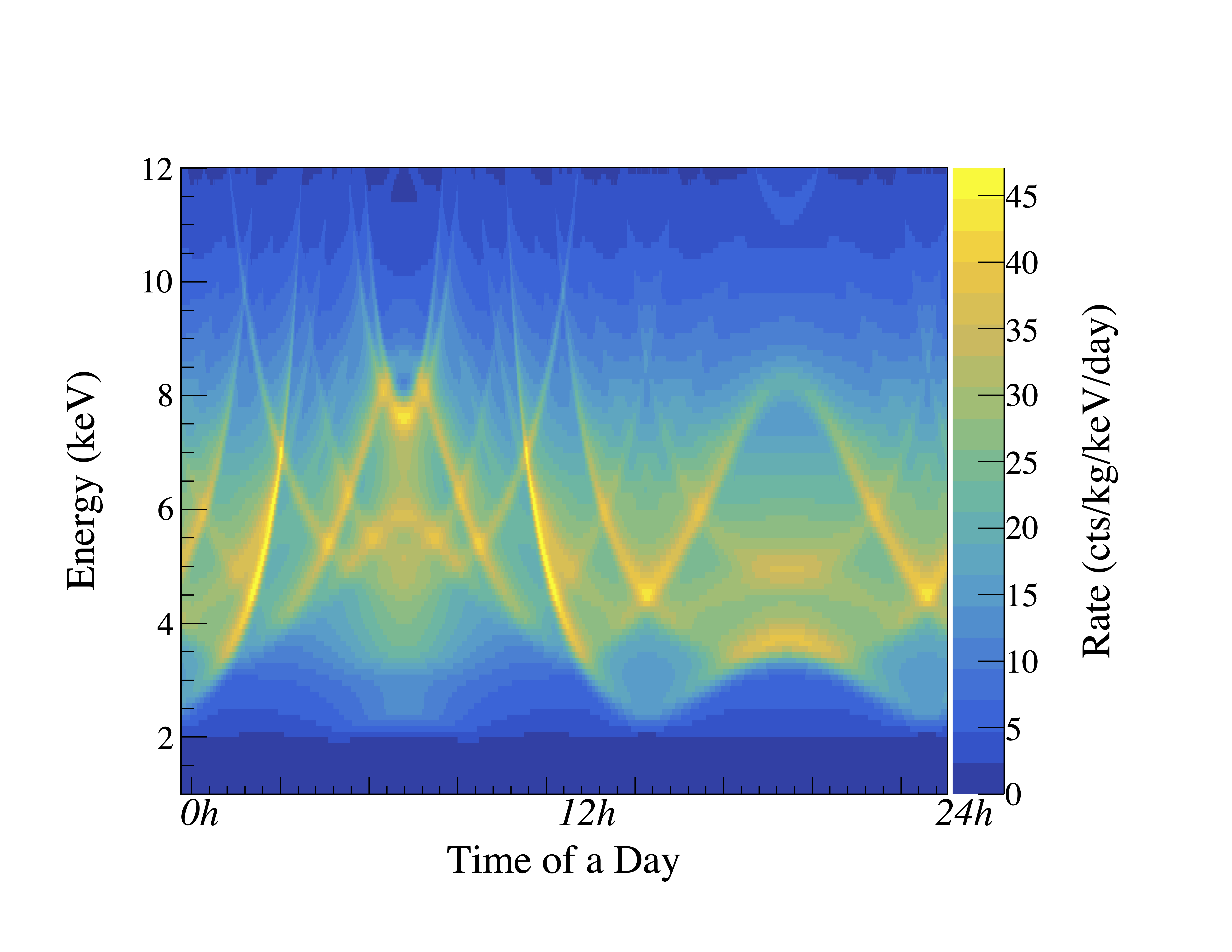}
    \caption{Axion signatures from coherent Primakoff-Bragg scattering averaged over all possible orientations of horizontal crystal planes, predicted for a 1-kg HPGe detector located at SURF for $g_{a\gamma}=10^{-8} $GeV$^{-1}$. The time duration shown is a full day.}
    \label{fig:average_pdf}
\end{figure}


If crystal horizontal orientations are unknown, the lack of information leads to reduced sensitivity, but the Primakoff-Bragg technique can still be exploited. 
The angle-dependent signal rate $R(\phi, E, t)$ can be averaged over all possible crystal horizontal orientations to give an angle-averaged signal rate $\bar R(E, t)$ in the entire experiment, as $\bar R(E, t)= \int^{\pi/4}_{-\pi/4} R(\phi, E, t)d\phi/(\pi/2)$. This angle averaging approach was first used for the DAMA best limit~\cite{Bernabei2001}. 
Within the Bayesian framework, the angle-averaging approach is the key step of marginalization over the nuisance parameter of crystal horizontal orientation. The prior for this nuisance parameter is uniform, if only the crystal boule axis is known. The angle-averaged signal rate still has strong time and energy dependence, as shown in Fig.~\ref{fig:average_pdf}, but it is not as distinct as those for specific crystal horizontal orientations, examples of which can be found in Refs.~\cite{Ahmed2009,Armengaud2013,Xu2017}. The reduction of sensitivity was found to be approximately a factor of 4 for $\lambda$~\cite{Xu2017}, which is about 40\% for \gag. 


The \MJD~experiment, located on the 4850' level at SURF~\cite{Heise2015}, searched for neutrinoless double beta (\onbb) decay in $^{76}$Ge with 44 kg of p-type, point-contact HPGe detectors, of which 30 kg were enriched in $^{76}$Ge~\cite{Abgrall2014,Alvis2019}.
\textsc{Majorana} detectors were deployed in two separate modules that were shielded against environmental radioactivity by layers of compact lead and copper shielding. 
The \DEM~ used extensive material screening and stringent cleanliness measures to achieve exceedingly low background. Surface exposure of the \enrge~ detectors was carefully limited, resulting in significantly lower background at low energy than the \natge~ detectors. The \DEM~ module 1 detectors started data taking in 2015, and module 2 became online in late 2016. Low energy physics was enabled by \DEM~ low-noise electronics~\cite{Abgrall2021}, and this analysis uses low-noise physics data taken between Jan. 2017 and Nov. 2019 by more than 20 \enrge~ detectors, with a background of about 0.05 counts/(keV kg d) at 5 keV. \textsc{Majorana} HPGe detectors have excellent energy resolution that is approaching 0.1$\%$ full-width at half maximum at the 2039 keV Q-value of the \onbb~decay~\cite{Alvis2019}, the best among all \onbb~ experiments. The 1-$\sigma$ energy resolution is better than 0.2 keV in the 5 to 22~keV energy range used in this analysis.  

\begin{figure}
    \includegraphics[width=8.3cm]{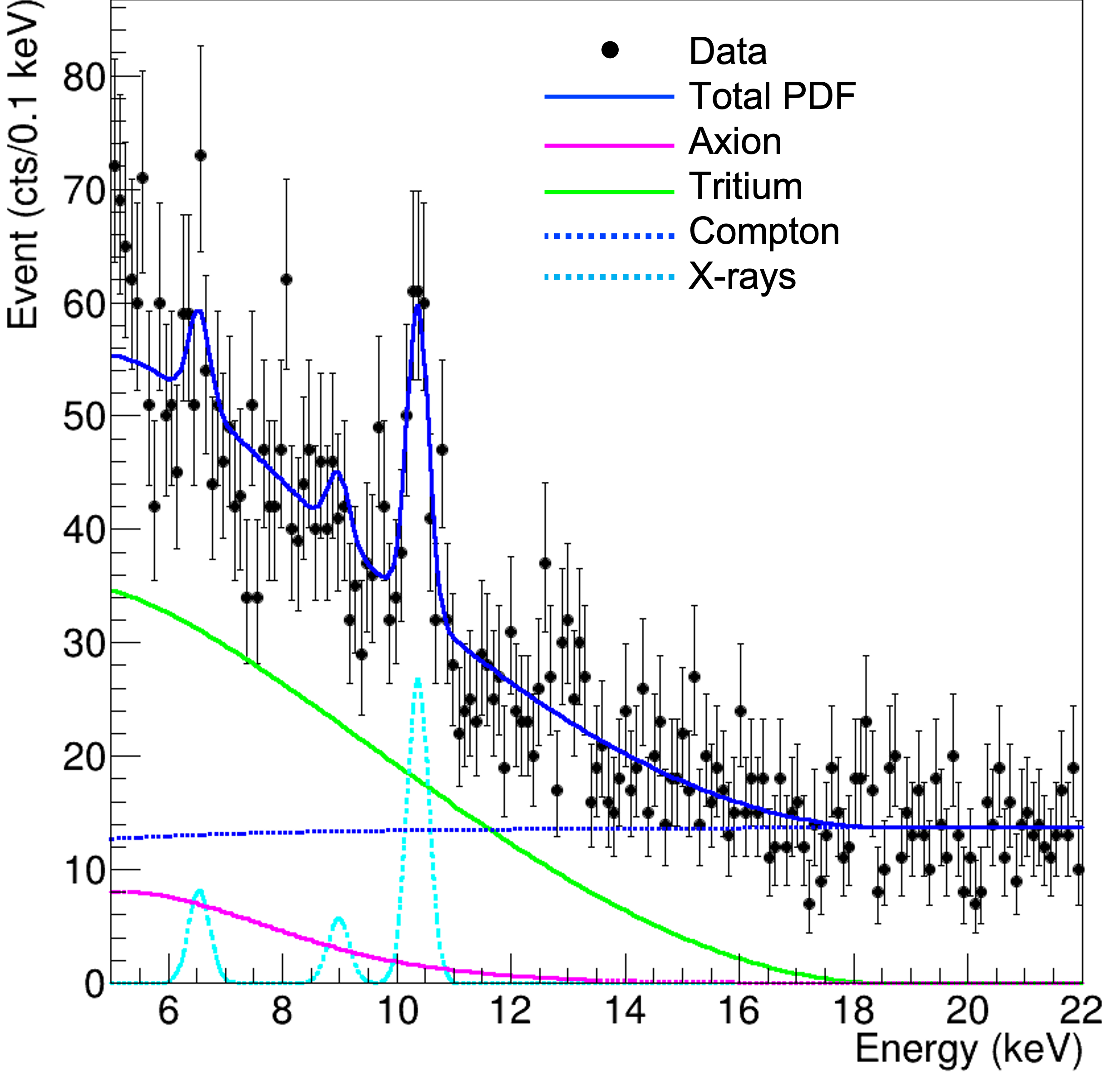}
    \caption{The energy spectrum of low energy events (Data), shown along with the best fit of the extended composite model (Total PDF) and individual components. All components have the energy efficiency folded.}
    \label{fig:ML_best_fit}
\end{figure}

HPGe detectors are known to have a transition layer immediately beneath the surface dead layer.~\cite{Aalseth2013}. Charge carriers diffuse out of the transition layer slowly, resulting in pulses with slower risetime and degraded energy. A novel data analysis method was recently developed to reject these slow pulses at low energy in the \DEM~\cite{Wiseman2018,Wiseman2020}. After denoising using a wavelet packet transform and a Daubechies-2 wavelet function, each low energy waveform is fitted with a heuristic exponentially modified Gaussian that generically resembles the waveform and features an effective slowness parameter corresponding to the slope
of the pulse rising edge. A cut is placed on the slowness parameter to accept fast waveforms only. The cut efficiency is calculated based on forward Compton scattering of 239-keV photons from routine $^{228}$Th source calibrations. By requiring the total energy to be shared in exactly two HPGe detectors, a sample dominated by real 239-keV photon events is obtained, which is made of bulk events since events with slow pulses do not have sum energy lying in the 239-keV peak due to energy degradation. Within this sample, detector hits with lower energy can be used to calculate the total cut efficiency, which is close to 90\% above 5 keV~\cite{Wiseman2022}. A paper with details on data selection, data cleaning, and analysis cuts used in low energy analysis of the \textsc{Demonstrator} is in preparation. To avoid lower efficiency and associated large relative uncertainties below 5 keV, this solar axion analysis starts at 5 keV and ends at 22 keV, which is beyond the 18.6-keV end point of tritium $\beta$-decay. The energy spectrum of the low energy events is shown in Fig~\ref{fig:ML_best_fit}.  

The data analysis has two steps. First, a maximum likelihood analysis was carried out to fit the data with a composite model with both axion signals and background components. Then, a Bayesian analysis was performed to extract an exclusion limit on axions. Due to lack of knowledge on the horizontal crystal plane angles for individual detectors, the angle-averaged Primakoff-Bragg axion signature $\bar R(E, t)$ is used to construct the signal model in both energy and temporal dimensions. The position of the Sun between Jan. 2017 and Nov. 2019 is tracked using the Naval Observatory Vector Astrometry Software (NOVAS)~\cite{USNO}, and the axion probability density function (PDF) is evaluated over the 3-year period with 5-minute precision. The background PDFs include the tritium $\beta$-decay, a flat background representing an extrapolation of the Compton continuum of photons at higher energy, and three Gaussian peaks representing the X-rays of cosmogenic $^{55}$Fe (6.54 keV), $^{65}$Zn (8.98 keV), and $^{68}$Ge (10.37 keV). Since tritium, $^{55}$Fe, $^{65}$Zn, and $^{68}$Ge were generated during limited surface exposure, they are modelled to decay away with corresponding half-lives, while the Compton continuum is constant in time. 
Both background and signal PDFs are weighted by the time-dependent exposure, which is the product of data-taking live time and active mass, and folded with the total cut efficiency in energy. A composite model is constructed by multiplying the PDF ($\mathcal{P}$) of the individual components with the corresponding strength ($n$) and adding them together, as expressed in Eq.~\ref{eq:model}, 
\begin{equation}
\mathcal{P}_{tot}(\vec x;\vec n) = n_{a}\mathcal{P}_\mathrm{a}(\vec x)+n_{C} \mathcal{P}_{C}(\vec x)  + n_T \mathcal{P}_T(\vec x) + \sum^{n_\mathrm{pks}} n_i\ \mathcal{P}_i(\vec x)  \label{eq:model}
 \end{equation}
  where $\vec x=(E, t)$ and $a, C, T, i$ indicate axions, Compton scatters, tritium, and the 3 X-ray peaks, respectively. The free parameters ($\vec n$) are the strengths of axion and backgrounds. Each individual PDF on the right side of Eq.~\ref{eq:model} is normalized to 1, so that the added total probability is normalized to the total number of events $n_D$ detected in the data. 
  \begin{equation}
     \int \mathcal{P}_{tot}(\vec x; \vec n)\ d\vec x = n_{a}+n_{C} + n_T+ \sum^{n_\mathrm{pks}} n_i\ = n_D\
  \end{equation}
Events from enriched detectors are combined together to form a single cohort and fitted with the composite model in an unbinned extended maximum likelihood method using the \texttt{RooFit} plus \texttt{RooStats} package~\cite{verkerke2006} in ROOT ver.6.22/06~\cite{Brun97}, the result of which is shown in Fig.~\ref{fig:ML_best_fit}. 
The number of axions is fitted to be $n_a$=297$\pm$138, which is consistent with zero at 2.2$\sigma$.

\begin{figure}
    \includegraphics[width=8.3cm]{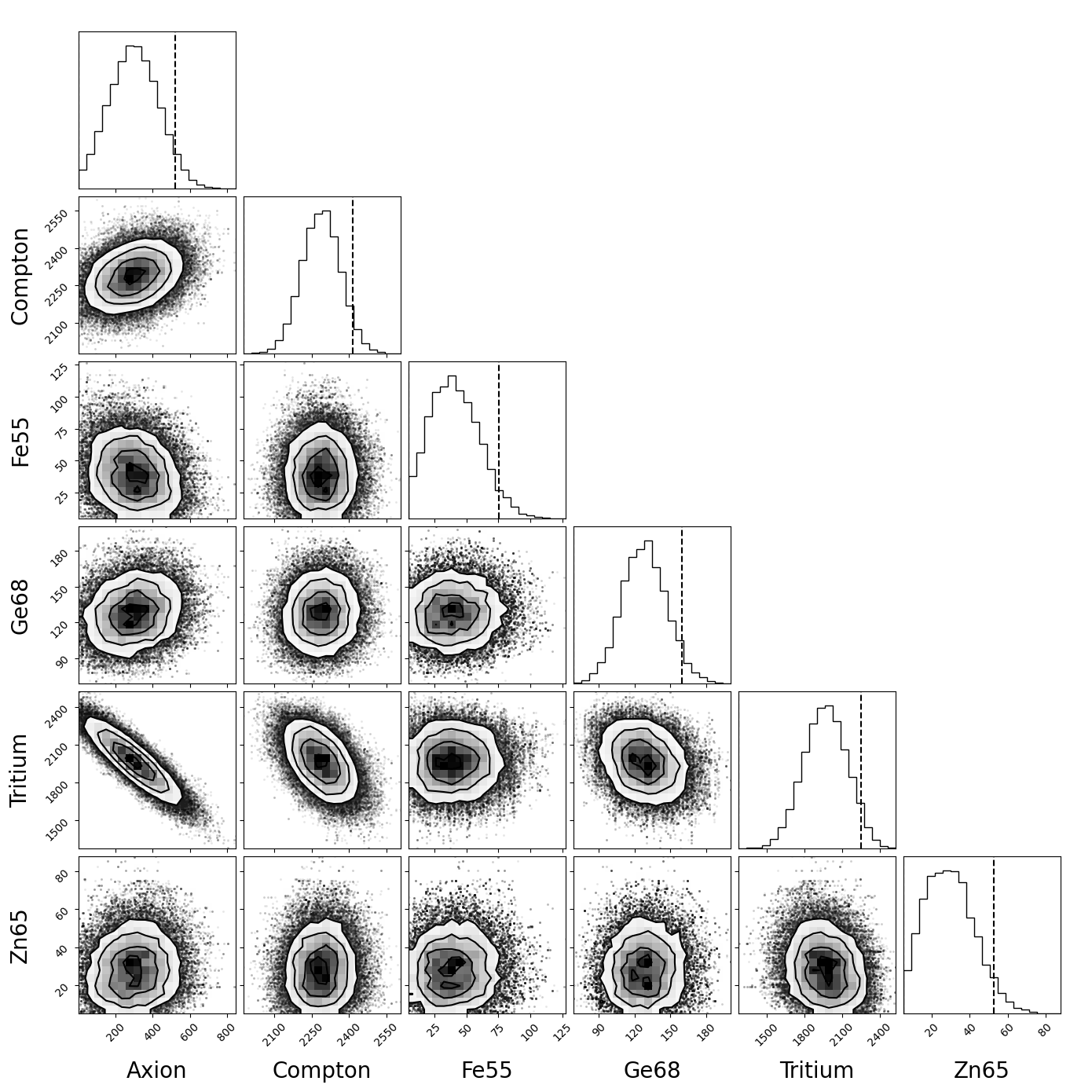}

    \caption{Contours for pairs of parameters in the \texttt{RooStats} MCMC sampling, and the projected posterior distributions for each parameter. Each vertical line in the posterior distributions indicates a 95\% interval.
    }
    \label{fig:axion_posterior}
\end{figure}

Ref.~\cite{Xu2017} pointed out that, when the angle averaged axion PDF is used, good statistical behaviors including the applicability of Wilks' theorem~\cite{Wilks1938} is guaranteed only for experiments with a dozen or more detectors. Otherwise, statistical sampling, such as Frequentist test statistic sampling as done by EDELWEISS~\cite{Armengaud2013} or Bayesian posterior sampling as in this analysis, will be necessary to correctly evaluate the statistical penalty of the information loss on the crystalline orientation. Since there are 20 or more detectors in this analysis, good statistical behaviors is guaranteed~\cite{Xu2017} and an estimation of Frequentist 95$\%$ limit on $n_a$ can be approximated as the best-fit plus 1.65$\sigma$ = 297$+1.65\times$138 = 525. A Bayesian analysis based on Markov-Chain Monte Carlo (MCMC) sampling was carried out for accurate limit setting. The Metropolis-Hastings (MH) sampling algorithm in \texttt{RooStats} is utilized to construct the MCMC chains with 5 million iterations. The axion number $n_a$ is the parameter of interest, and the rest of $\vec n$ are nuisance parameters. The other nuisance parameter $\phi$ was already marginalized in $\bar R(E,t)$. A uniform prior between 0 and all events in the analyzed datasets is used for $n_a$, similar to the CDMS analysis~\cite{Ahmed2009}. Posterior probability distributions after sampling are shown in Fig.~\ref{fig:axion_posterior}, along with correlations among parameters. The median values in the Bayesian posteriors match well with their corresponding maximum-likelihood best fits. $n_a$ is found to be less than 519 at a 95\% CL. This 95\% Bayesian upper limit on $n_a$ is close to the approximate Frequentist limit and it translates to $\lambda< 4.48\times10^{-4}$ or $g_{a\gamma}<1.45\times 10^{-9}$ GeV$^{-1}$, improving on the limits from DAMA (90\% CL) and EDELWEISS-II (95\% CL). 


Systematic uncertainties arise from cut efficiency and energy.  The dominant contribution is cut efficiency, where lower efficiency leads to a worse limit and vice-versa. By varying the cut efficiency within its own uncertainties~\cite{Wiseman2022}, a change of up to 16\% is found on the limit of $\lambda$. Based on the cosmogenic X-ray peaks at low energy, the energy scale is found to be accurate to within 0.1 keV of the expected values~\cite{Kim2022,Wiseman2022}. An energy scale uncertainty of $\pm0.1$~keV introduces a 6\% uncertainty in the limit. The variation of energy resolution throughout time yields another 5\% uncertainty. In total, an 18\% systematic uncertainty is associated with the limit on $\lambda$, which is a 5\% uncertainty on \gag.

\begin{figure}
    \includegraphics[width=8.3cm]{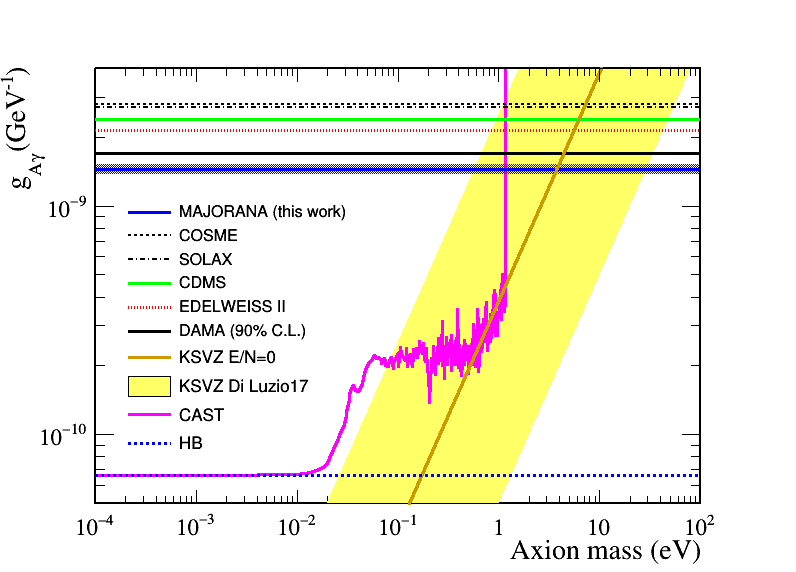}
    \caption{\gag~limit obtained in this analysis (solid blue line), with the uncertainty shown as a band around the limit. The HB constraints~\cite{Ayala2014}, the CAST limits~\cite{Andriamonje2007,Arik2011, Arik2015, Anastassopoulos2017}, and other solid state limits are also shown for SOLAX~\cite{Avignone1998}, DAMA~\cite{Bernabei2001}, COSME~\cite{Morales2002}, CDMS~\cite{Ahmed2009} and EDELWEISS-II~\cite{Armengaud2013}. All limits are for 95\% CL, except for the 90\% DAMA limit. The KSVZ axion phase space is shown with the realistic range of $E/N$ found in~\cite{DiLuzio2017}.  }
    \label{fig:gagg}
\end{figure}

In conclusion, a temporal-energy analysis was performed by the \MJD~experiment to search for coherent Primakoff-Bragg axion signals between 5 and 22 keV with a total exposure of 33 kg-yr, the rate of which is found to be consistent with zero. A new limit of 95\% CL on the axion-photon coupling is obtained as \gag$<1.45\times 10^{-9}$ GeV$^{-1}$ in a Bayesian approach. This improves the limit in laboratory searches for axion mass between 1.2 eV/$c^2$ and 100 eV/$c^2$, as shown in Fig.~\ref{fig:gagg}, and it continues to exclude KSVZ axions in the larger phase space. An intriguing excess of events was observed by the XENON1T experiment at 1-5 keV~\cite{Aprile2020}, which could be due to tritium background or a number of new physics signals. Examining the XENON1T excess against the solar axion hypothesis leads to a range of theoretical considerations~\cite{DiLuzio2020, Dent2020, Gao2020}. The limit established in this work is not sensitive enough to exclude the possibility of the XENON1T excess being Primakoff solar axions.

This work presented here by the \textsc{Demonstrator} is one important step forward to improve the best limit in a long series of experimental efforts of using solid-state detectors to effectively probe the 1 eV/$c^2$ to 100 eV/$c^2$ mass range, more than 20 years after the DAMA best limit was published. The analysis method used here can be readily employed by the ton-scale Large Enriched Germanium Experiment for Neutrinoless $\beta\beta$ Decay (LEGEND)~\cite{LEGEND21}. With a 10 ton-year exposure and lower background than the \textsc{Demonstrator}, LEGEND has the potential to effectively probe \gag~well below $10^{-9}$ GeV$^{-1}$ without measuring the crystal orientations of hundreds of individual detectors. 

  \begin{acknowledgments}
  This material is based upon work supported by the U.S.~Department of Energy, Office of Science, Office of Nuclear Physics under contract / award numbers DE-AC02-05CH11231, DE-AC05-00OR22725, DE-AC05-76RL0130, DE-FG02-97ER41020, DE-FG02-97ER41033, DE-FG02-97ER41041, DE-SC0012612, DE-SC0014445, DE-SC0018060, and LANLEM77/LANLEM78. We acknowledge support from the Particle Astrophysics Program and Nuclear Physics Program of the National Science Foundation through grant numbers MRI-0923142, PHY-1003399, PHY-1102292, PHY-1206314, PHY-1614611, PHY-1812409, PHY-1812356, and PHY-2111140. We gratefully acknowledge the support of the Laboratory Directed Research \& Development (LDRD) program at Lawrence Berkeley National Laboratory for this work. We gratefully acknowledge the support of the U.S.~Department of Energy through the Los Alamos National Laboratory LDRD Program and through the Pacific Northwest National Laboratory LDRD Program for this work.  We gratefully acknowledge the support of the South Dakota Board of Regents Competitive Research Grant. We acknowledge the support of the Natural Sciences and Engineering Research Council of Canada, funding reference number SAPIN-2017-00023, and from the Canada Foundation for Innovation John R.~Evans Leaders Fund.  This research used resources provided by the Oak Ridge Leadership Computing Facility at Oak Ridge National Laboratory and by the National Energy Research Scientific Computing Center, a U.S.~Department of Energy Office of Science User Facility located at Lawrence Berkeley National Laboratory. We thank our hosts and colleagues at the Sanford Underground Research Facility for their support.
    \end{acknowledgments}
\bibliography{main}

\end{document}